\documentclass[twocolumn]{revtex4-1}
\usepackage{graphicx} 
\usepackage{dcolumn} 
\usepackage{bm} 
\usepackage{amssymb} 
\usepackage{latexsym} 
\usepackage{tensor} 
\begin{document} 
\title{Correspondences between scalar field  and fluid fluctuations in curved
spacetime}
\author{ Seema Satin }
\affiliation{Indian Institute for Science Education and Research, Kolkata, 
India}
\email{seemasatin74@gmail.com} 
\newcommand{\tG}{\tensor{G}} 
\newcommand{\hT}{\hat{T}}
\newcommand{\C}{\mbox{Cov}} 
\newcommand{\be}{\begin{equation}} 
\newcommand{\ee}{\end{equation}} 
\newcommand{\bea}{\begin{eqnarray}} 
\newcommand{\eea}{\end{eqnarray}}
\newcommand{\G}{\hat{G}} 
\newcommand{\e}{\epsilon} 
\begin{abstract} 
A correspondence between scalar field fluctuations and generalized 
fluctuations in a hydrodynamic approximation of fields is obtained. 
The results presented here are of interest to field-fluid correspondences and
form part of theoretical foundations in this area.  The intention for
such developments is to explore sub-hydro range mesoscopic physics for the 
relativistic fluids in curved spacetime. 
The fluid correspondences fall in the classical domain and  can replace the
 quantum fields and fluctuations for scales around the hydrodynamic limits. 
The present article extends our earlier results with a more elaborate physical
 insight towards the quantum fluids and retention of partial quantum nature
in a stochastic description in bulk of the fluids.  
This also accounts for  non-thermal effects along with  thermal and  quantum 
fluctuations for the fields in the hydro limit. Hence the expressions
 presented here are very general in nature for various applications.
The further scope of research  that such developments give is discussed in
the concluding section. 
\end{abstract}
\maketitle
\section{Introduction}
A correspondence between the stress energy tensor for a scalar field and that
 of a fluid is well known and widely accepted \cite{madsen,mainini} . The 
 field-fluid correspondence is of interest in various aspects and is an active
 area of research \cite{mukund,valerio,semiz}.    
In this article, we show the correspondence between fluctuations of
quantum fields and relativistic fluids in a hydrodynamic description. 
The results presented in this article are of significance for quantum fluids
on a spacetime structure, where the quantum nature in the bulk  persists
in the macroscopic hydro-approximation.
This is  physically a different situation than that considered in 
\cite{seemacorr, seemacqg}
where the approximation is taken after de-coherence of quantum fields
takes place. This difference is typically characterised in this article by
giving a stochastic nature  to the the kinetic term (four-velocity),  
 along with the other fluid variables, whereas the four-velocity 
 is considered as a deterministic variable in \cite{seemacorr,seemacqg}
for the decoherence limit correspondences.  The expressions that we obtain here
are full results in the sense of the field-fluid correspondence and can be
used for various applications in relativistic astrophysics. Thus a mesoscopic
theory for relativistic fluids with the inclusion of fluctuations in the 
hydrodynamic approximation from the first principles can be formulated.
There are ongoing efforts for relating  thermal fluctuations
to microscopic details  in relativistic fluids \cite{kovtun},
however we present a different formulation with our correspondence. In our 
approach the
quantum field fluctuations lead to the hydro-variable fluctuations
in a coarse grained fashion, without the need to necessarily relate
 thermodynamic
details in the fluids. Thus the expressions that we work out in this article,
 hold even for non-thermal effects and hence have a fundamentally 
different origin.  This also provied  a wider span of applications and
 inspection than the other approaches based on thermal fluids.

We carry  out  our work by relating the fluctuations in terms of a two point
 noise kernel for the field and fluid approximations. For scalar fields
such a noise kernel has been elaborately obtained in  \cite{bei1}. In this
 article, we consider the case of a quantum fluid in
 which the kinetic term as well as the bulk variables show corresponding
 random nature. 
One can surely, also address thermal fluctuations in fluids  with such a 
prescription.  Hence, this is of importance  as a mesoscopic
approach to probe intermediate scales in  quantum fluids, and is very general
w.r.t origin of various kinds of stochastic effects.

On the other hand, for the theory of semiclassical stochastic gravity 
\cite{phillipnk}, the point separated semiclassical noise kernel is of
central importance to study structure formation in the early universe 
\cite{bei2}. 
A similar feature in order to probe  the sub-hydro mesoscopic scales
 can be of interest for a different (later) epoch in the
evolution of the universe.  Our developments given here can also be
of significance  for the exotic matter fields that the relativistic stars are
 composed of. This opens up a new direction to explore in the area of 
 massive stars and relativistic fluids.  Current efforts towards
 relativistic fluids for the massive star interiors are being  persued
 \cite{nils,schmidt} and gaining more attention with the
 achievements in gravitational waves \cite{abbot,abbot1,nergis} detection. 
The study of fluctuations of dense matter fluids is hence also of significance
 to asteroseismology. 

Our framework  is that of building up a theoretical base to address the
new intermediate scales for study in the above 
two  areas of application, along with interest in the field-fluid 
correspondence, which is the focus in this article. 

For modelling the hydro-limit of fields one has to  consider the classical 
relativisitic fluid approximation in terms of  the stress tensors. 
The spacetime metric $ g_{ab}$ on which a relativistic fluid exists 
is considered to be deterministic in our work. 
The generalized stochasticity concept (\cite{seemacqg})takes into consideration 
"roughness in physical variables" at mesoscopic scales which are yet 
unexplored in relativistic fluids. We begin with quantum fields and their
fluctuations in terms of the semiclassical noise kernel.   
\section{Fluctuations of quantum fields in curved spacetime}
In this section we revise the two point semiclassical noise kernel, which 
defines the fluctuations of quantum scalar fields.
Later in this article, we will show its relation and correspondence with the
 hydro approximation. 

The stress energy tensor for quantum fields is given by 
\begin{widetext}
\be \label{eq:1}
\hat{T}^{(field)}_{ab}(x)  =  \phi_{;a} \phi_{;b} - \frac{1}{2} g_{ab}
( \phi^{;c} \phi_{;c}  +  m^2 \phi^2 )
+ \xi (g_{ab} \Box - \nabla_a \nabla_b + G_{ab} )\phi^2
\ee
\end{widetext}
Fluctuations in quantum fields are defined by  the bitensor
which is  a two point noise kernel  $N_{abcd}(x,x')$  as obtained  in the
 theoretical developments of semiclassical stochastic gravity \cite{bei1}.
\be \label{eq:nfluc}
\scalebox{.8}{ $ 8 N_{abc'd'} (x,x')  =  <\{ \hT_{ab}(x),\hT_{c'd'}(x') \}> -
2 <\hT_{ab}(x)><\hT_{c'd'}(x')> $ }
\ee
where $<...>$ denotes the  expectation of the quantum field $\phi$ on the 
spacetime background ( in this article we denote the quantum expectation
with $<....>$ and classical averages or expectation with $E(...) $ ),
where
\begin{widetext}
\be  \label{eq:n12}
N_{abc'd'}(x,x') = \tilde{N}_{abc'd'}(x,x') + g_{ab}(x) 
\tilde{N}_{c'd'}(x,x')
 + g_{c'd'}(x') \tilde{N'}_{ab}(x,x') + g_{ab}(x) g_{c'd'}(x') \tilde{N}(x,x')
\ee
Further on  we prescribe the noise kernel expressions for the  field as
 $N^{(field)}_{abc'd'} $ and for fluid as $N^{(fluid)}_{abc'd'} $. 
For the scalar fields then, with non-minimal coupling, the noise kernel is
 given as 
\bea \label{eq:N2}
 8 \tilde{N}^{(field)}_{abc'd'} & = & (1-2 \xi)^2 (G_{;c'b} G_{;d'a} + G_{;c'a}
 G_{;d'b} ) + 4 \xi^2 ( G_{;c'd'} G_{;ab} + G G_{;abc'd'} ) - 2 \xi (1-2 \xi)
 (G_{;b}G_{;c'ad'}+ G_{;a} G_{;c'bd'} \nonumber \\
& &  + G_{;d'} G_{;abc'} + G_{;c'} G_{;abd'} ) + 2 \xi (1-2 \xi)
( G_{;a}G_{;b} R_{c'd'} + G_{;c'}G_{;d'} R_{ab} ) - 4 \xi^2 ( G_{;ab} R_{c'd'} + G_{;c'd'} R_{ab} )
 G \nonumber \\
& &  + 2 \xi^2 R_{c'd'} R_{ab} G^2 \\
\label{eq:N3} 8 \tilde{N}^{'(field)}_{ab} & =  &
 2 ( 1- 2\xi)[( 2 \xi- \frac{1}{2} ) G_{;p'b}
\tG{_;^{p'}_a }+ \xi (G_{;b} \tG{_;_{p'}_a^{p'}} + G{;_a} \tG{_;_{p'}_b^{p'}})]
 - 4 \xi [(2 \xi - \frac{1}{2} ) \tG{_;^{p'}} \tG{_;_a_b_{p'}} \nonumber \\
& & + \xi ( \tG{_;_{p'}^{p'}} \tG{_;_a_b} + G \tG{_;_a_b_{p'}^{p'}} ) ]
 -(m^2 + \xi R') [ ( 1 - 2 \xi ) \tG{_;_a} \tG{_;_b} - 2 G \xi \tG{_;_a_b} ]
+ 2 \xi [ ( 2 \xi - \frac{1}{2}) G_{; {p'}} \tG{_;^{p'}} + \nonumber \\
& &  2 G \xi \tG{_;_{p'}^{ p'}} ] R_{ab} - (m^2 + \xi R' ) \xi R_{ab} G^2 \\
\label{eq:N4} 8 \tilde{N}^{(field)}  & = & 2 ( 2 \xi - \frac{1}{2})^2 
\tG{_;_{p'}_q}
\tG{_;^{p'}^{q}} + 4 \xi^2 ( \tG{_;_{p'}^{p'} } \tG{_;_q^q}
 + G \tG{_;_p^p_{q'}^{q'}} ) + 4 \xi ( 2 \xi - \frac{1}{2} ) ( G_{;p}
\tG{_;_{q'}^p^{q'}}  + \tG{_;^{p'}} \tG{_;_q^q_{p'}})   \nonumber \\
& & - (2 \xi - \frac{1}{2} ) [ (m^2 + \xi R )G_{;p'} \tG{_;^{p'}} + (m^2 + \xi
R') G_{;p} \tG{_;^p}]
 - 2 \xi [ (m^2 + \xi R) \tG{_;_{p'}^{p'}} + (m^2 + \xi R')
\tG{_;_p^p}] G \nonumber \\
& &   + \frac{1}{2} (m^2 + \xi R ) ( m^2 + \xi R') G^2 \\
8 \tilde{N}^{(field)}_{c'd'} & = &  2 (1-2 \xi)^2 \tG{_;_{c'}_p}
\tG{_;_{d'}^p} +
 4 \xi^2 ( \tG{_;_{c'}_{d'}} \Box_x G + G \Box_x \tG{_;_{c'}_{d'}}) -
 2 \xi (1-2\xi) (2 \tG{_;_p} \tG{_;_{c'}^p_{d'}}  \nonumber \\
& & + \tG{_;_{d'}} \Box_x \tG{_;_{c'}} + \tG{_;_{c'}} \Box_x
\tG{_;_{d'}} ) + 2 \xi ( 1-2 \xi) ( \tG{_;_p} \tG{_;^p} R_{c' d'} + \tG{_;_{c'}}
\tG{_;_{d'}} R ) - 4 \xi^2 ( \Box_x G R_{c'd'} \nonumber \\
& &  + \tG{_;_{c'}_{d'}} R ) G + 2 \xi^2 R_{c' d' } R G^2 
\eea
\end{widetext}
where $ G  \equiv G(x,x') $ are the Wightman functions defined by
$<\phi(x) \phi(x') > $. 
Our aim is to show a correspondence of these fluctuations with
the sub-hydro limit fluctuations. These sub-hydro scales that we intend
to address here, are expected to lie a little below the classical macroscopic
 hydrodynamic scales and much above the quantum microscopic scales.   
We will take the classical limit
of the stress tensor for the fluid approximation and show the explicit form
of the  fluctuations that give  access to a new regime between
macro and micro scales in a straightforward way.   
\section{Generalized fluctuations in the fluid approximation and their 
correspondence with the field fluctuations}
The quantum fields can be treated as a fluid in the hydrodynamic approximation,
such that, the fluid variables associated with the fields are given by 
\cite{madsen},
\begin{widetext}
\bea
u_a & = &  [\partial_c (\phi) \partial^c (\phi) ]^{-1/2} \partial_a
\phi  \label{eq:vel}\\
\epsilon & =& (1- \xi \phi^2)^{-1}[ \frac{1}{2} \partial_c \phi \partial^c
 \phi + V(\phi) + \xi \{\Box(\phi^2) - (\partial^c \phi \partial_c \phi )^{-1}
\partial^a \phi \partial^b \phi \nabla_a \nabla_b ( \phi^2) \}] \label{eq:ep}\\
q_a &= & \xi ( 1- \xi \phi^2)^{-1} ( \partial^b \phi \partial_b \phi ) ^{-3/2}
\partial^c \phi \partial^d \phi [ \nabla_c \nabla_d (\phi^2) \partial_a \phi
- \nabla_a \nabla_c (\phi^2) \partial_d \phi ] \label{eq:q}\\
p & = & (1- \xi \phi^2)^{-1} [ \frac{1}{2} \partial_c  \phi \partial^c
 \phi - V(\phi) -  \xi \{ \frac{2}{3} \Box (\phi^2) + \frac{1}{3} (\partial_c
 \phi \partial^c \phi )^{-1} \nabla_a \nabla_b (\phi^2) \partial^a \phi
\partial^b \phi \} ] \label{eq:p} \\
\pi_{ab} & = & \xi (1- \xi \phi^2 )^{-1} ( \partial^c \phi \partial_c \phi)^{
-1} [
\frac{1}{3} ( \partial_a \phi \partial_b \phi - g_{ab} \partial^c \phi
 \partial_c \phi ) \{ \Box ( \phi^2) - (\partial^c \phi \partial_c \phi )^{-1}
\nabla_c \nabla_d (\phi^2) \partial^c \phi \partial^d \phi \} \nonumber \\
& &   + \partial^p \phi \{ \nabla_a \nabla_b ( \phi^2) \partial_p \phi -
\nabla_a \nabla_p (\phi^2) \partial_b \phi - \nabla_p \nabla_b
( \phi^2) \partial_a \phi + ( \partial_c \phi \partial^c \phi ) ^{-1}
\partial^d \phi \nabla_d \nabla_p (\phi^2) \partial_a \phi \partial_b \phi \}]
\label{eq:pi}
\nabla_a \nabla_p (\phi^2) \partial_b \phi  \nonumber \\
& &- \nabla_p \nabla_b ( \phi^2) \partial_a \phi + ( \partial_c \phi 
\partial^c \phi ) ^{-1} \partial^d \phi \nabla_d \nabla_p (\phi^2) 
\partial_a \phi \partial_b \phi \}]
\label{eq:pi}
\eea
where  the fluid stress tensor is of the form,
\be
T^{(fluid)}_{ab} = u_a u_b(\e+p) + g_{ab} p + q_a u_b + u_a q_b + \pi_{ab}
\ee
The variables $\epsilon, p, u_a, q_a , \pi_{ab} $, denote the energy density 
, pressure, four-velocity, heat flux and anisotropic stresses respectively
in the fluid.
The noise kernel in  equation (\ref{eq:nfluc}) in the hydrodynamic
 approximation then takes an overall
classical form in terms of the fluid  stress tensor, 
\be
8 N^{(fluid)}_{abc'd'}(x,x') = 2(E( T^{(fluid)}_{ab}(x) 
T^{(fluid)}_{c'd'}(x')) - E(T^{(fluid)}_{ab}(x)) E( T^{(fluid)}_{c'd'}(x')) 
= 2 \C[T^{(fluid)}_{ab}(x), T^{(fluid)}_{c'd'}(x')] 
\ee
where $\C$ represents covariance. 
The two point covariance for the  stress tensor 
$ \C[T^{(fluid)}_{ab}(x), T^{(fluid)}_{c'd'}(x')] $ can be worked out 
easily, and terms arranged in
 order such that,  coefficients of the metric $g_{ab}$,  as in equation
 (\ref{eq:n12}) can be shown to correspond to the noise kernel for the fields
 as,  
\bea 
 & & \tilde{N}^{(field)}_{abc'd'} \Longrightarrow \frac{1}{2}
\tilde{N}^{(fluid)}_{abc'd'} (\mbox{ relating the terms as : } ) \nonumber 
\\
  &  & (1-2 \xi)^2 (G_{;c'b} G_{;d'a} + G_{;c'a}
 G_{;d'b} ) + 4 \xi^2 ( G_{;c'd'} G_{;ab} + G G_{;abc'd'} ) - 2 \xi (1-2 \xi)
 (G_{;b}G_{;c'ad'}+ G_{;a} G_{;c'bd'} \nonumber \\
& &  + G_{;d'} G_{;abc'} + G_{;c'} G_{;abd'} ) + 2 \xi (1-2 \xi)
( G_{;a}G_{;b} R_{c'd'} + G_{;c'}G_{;d'} R_{ab} ) - 4 \xi^2 ( G_{;ab} R_{c'd'} + G_{;c'd'} R_{ab} )
 G \nonumber \\
& &  + 2 \xi^2 R_{c'd'} R_{ab} G^2 \Longrightarrow 
\frac{1}{2}\{\{ E(\e(x)) E(\e(x')) + E(\e(x)) E(p(x')) + E(p(x)) E(\e(x')) +
 E(p(x)) E(p(x')) \} \nonumber \\
& & [4 E(u_{(a}) E(u_{(c'}) \C[u_{b)}, u_{d')}] 
+ 2 \C[u_a, u_{(c'}] \C[u_b, u_{d')}] ] + \{ \C[\e(x),\e(x')] + 
 \C[\e(x), p(x')] + \C[p(x), \e(x')] \nonumber \\
& &  + \C[p(x), p(x')] \} \{ E(u_a) E(u_b) E(u_{c'}) E(u_{d'}) + 
 4  E(u_{(a}) E(u_{(c'}) \C[u_{b)}, u_{d')}] + 2\C[u_a, u_{(c'}]
 \C[u_b, u_{d')}]\}  +  \nonumber \\
& & 4\{ (E(p(x)) +  E(\e(x)))  E(u_{(a}) E(q_{(c'}) +  
  (E(p(x')) + E(\e(x')))  E(q_{(a}) E(u_{(c'})\} \C[u_{b)}, u_{d')}] + 
  4 E(u_{(a}) E(u_{(c'}) \C[q_{b)}, q_{d')}]  \nonumber \\
& & + \C[q_a, q_{d'}] \C[u_b, u_{c'}] + 2 E(q_a) E(q_{(c'}) \C[u_a, u_{d')}]
 + \C[\pi_{ab}, \pi_{c'd'} ] \} 
\eea
 The notation
for terms like $(u_{(a} , q_{b)}) $ represents $\frac{1}{2} ( u_{a} q_b + 
q_a u_b)$. Also note that the indices $(a,b)$ and $(c',d')$ are not to be
 mixed up in the above expressions for the two points $x$ and $x'$, as  $a,b$ 
remain  
associated with $x$, while $c',d'$ remain associated with $x'$ for all the
terms above and the following expressions as well. 

Coefficient of $g_{ab}$ 
\bea
& &  \tilde{N}^{'(field)}_{c'd'} \Longrightarrow \frac{1}{2}
\tilde{N}^{'(fluid)}_{cd} (\mbox{ relating the terms as : } )  \nonumber \\
&  &  2 (1-2 \xi)^2 \tG{_;_{c'}_p} \tG{_;_{d'}^p} +
 4 \xi^2 ( \tG{_;_{c'}_{d'}} \Box_x G + G \Box_x \tG{_;_{c'}_{d'}}) -
 2 \xi (1-2\xi) (2 \tG{_;_p} \tG{_;_{c'}^p_{d'}}  \nonumber \\
& & + \tG{_;_{d'}} \Box_x \tG{_;_{c'}} + \tG{_;_{c'}} \Box_x
\tG{_;_{d'}} ) + 2 \xi ( 1-2 \xi) ( \tG{_;_p} \tG{_;^p} R_{c' d'} + \tG{_;_{c'}}
\tG{_;_{d'}} R ) - 4 \xi^2 ( \Box_x G R_{c'd'} \nonumber \\
& &  + \tG{_;_{c'}_{d'}} R ) G + 2 \xi^2 R_{c' d' } R G^2  
\Longrightarrow
 E(u_{c'}) E(u_{d'}) \{ \C[p(x), p(x')]+ \C[p(x),\e(x')] \} 
\eea
Coefficient of $g_{c'd'}$,
\bea
 & & \tilde{N}^{(field)}_{ab} \Longrightarrow \tilde{N}^{(fluid)}_{ab} (\mbox{
relating the terms as : }) \nonumber \\
&  & 2 ( 1- 2\xi)[( 2 \xi- \frac{1}{2} ) G_{;p'b}
\tG{_;^{p'}_a }+ \xi (G_{;b} \tG{_;_{p'}_a^{p'}} + G{;_a} \tG{_;_{p'}_b^{p'}})]
 - 4 \xi [(2 \xi - \frac{1}{2} ) \tG{_;^{p'}} \tG{_;_a_b_{p'}} \nonumber \\
& & + \xi ( \tG{_;_{p'}^{p'}} \tG{_;_a_b} + G \tG{_;_a_b_{p'}^{p'}} ) ]
 -(m^2 + \xi R') [ ( 1 - 2 \xi ) \tG{_;_a} \tG{_;_b} - 2 G \xi \tG{_;_a_b} ]
+ 2 \xi [ ( 2 \xi - \frac{1}{2}) G_{; {p'}} \tG{_;^{p'}} + \nonumber \\
& &  2 G \xi \tG{_;_{p'}^{ p'}} ] R_{ab} - (m^2 + \xi R' ) \xi R_{ab} G^2 
\Longrightarrow
\frac{1}{2}\{ E(u_a) E(u_b) \{ \C[\e(x),p(x')] + \C[p(x),p(x')] \} \} 
\eea
Coefficient of $g_{ab} g_{c'd'} $,
\bea
& &  \tilde{N}^{(field)} \Longrightarrow \frac{1}{2}\tilde{N}^{(fluid)} 
(\mbox{relating the terms as: })  \nonumber \\
&  & 2 ( 2 \xi - \frac{1}{2})^2 \tG{_;_{p'}_q}
\tG{_;^{p'}^{q}} + 4 \xi^2 ( \tG{_;_{p'}^{p'} } \tG{_;_q^q}
 + G \tG{_;_p^p_{q'}^{q'}} ) + 4 \xi ( 2 \xi - \frac{1}{2} ) ( G_{;p}
\tG{_;_{q'}^p^{q'}}  + \tG{_;^{p'}} \tG{_;_q^q_{p'}})   \nonumber \\
& & - (2 \xi - \frac{1}{2} ) [ (m^2 + \xi R )G_{;p'} \tG{_;^{p'}} + (m^2 + \xi
R') G_{;p} \tG{_;^p}] - 2 \xi [ (m^2 + \xi R) \tG{_;_{p'}^{p'}} + 
(m^2 + \xi R') \tG{_;_p^p}] G \nonumber \\
& &   + \frac{1}{2} (m^2 + \xi R ) ( m^2 + \xi R') G^2 
\Longrightarrow
 \frac{1}{2} \C[p(x) p(x')] 
\eea
\end{widetext}
In the above, we can see that the two point covariances of the heat flux, 
anisotropic stesses, as well as for the four-velocity appear only in the 
part $\tilde{N}_{abc'd'} $ for the noise kernel of the fluid.  For the
rest of parts, the two point covariances of the pressure and energy
density suffice to characterize them.  Given the complex nature of the
above expressions, it is operationally difficult to obtain reverse
equations  with one  to one correspondence between each of the the two point
fluid variables and that of the fields.  
\subsection{Perfect fluid case }
For the perfect fluid case, where  the heat flux $q_a$ and  anisotropic 
stresses $\pi_{ab} $ vanish , which can be related to the vanishing
of the the non-minimal coupling factor $\xi$, for the scalar field
stress tensor, one has 
\bea
u_a & = &  [\partial_c (\phi) \partial^c (\phi) ]^{-1/2} \partial_a
\phi  \label{eq:vel1} \nonumber \\
\epsilon & =& [ \frac{1}{2} \partial_c \phi \partial^c
 \phi + V(\phi)  \label{eq:ep1} \nonumber \\
p & = &  [ \frac{1}{2} \partial_c  \phi \partial^c
 \phi - V(\phi) ] \label{eq:p1} 
\eea
and  the noise kernel parts are related by, 
\begin{widetext}
\bea 
 & &   \tilde{N}^{(field)}_{abc'd'} \longrightarrow 
\frac{1}{2} \tilde{N}^{(fluid)}_{abcd}
(\mbox{(relating the corresponding term as :)}  \nonumber \\ 
& &  \{ G_{;c'b} G_{;d'a} + G_{;c'a} G_{;d'b} \}
\longrightarrow
\frac{1}{2}\{\{ E(\e(x)) E(\e(x')) + E(\e(x)) E(p(x')) + E(p(x)) E(\e(x')) +
 E(p(x)) E(p(x')) \} \nonumber \\
& & [4 E(u_{(a}) E(u_{(c'}) \C[u_{b)}, u_{d')}] 
+ 2 \C[u_a, u_{(c'}] \C[u_b, u_{d')}] ] + \{ \C[\e(x),\e(x')] + 
 \C[\e(x), p(x')] + \C[p(x), \e(x')] \nonumber \\
& &  + \C[p(x), p(x')] \} \{ E(u_a) E(u_b) E(u_{c'}) E(u_{d'}) + 
 4  E(u_{(a}) E(u_{(c'}) \C[u_{b)}, u_{d')}] + 2\C[u_a, u_{(c'}]
 \C[u_b, u_{d')}]\}  +  \nonumber \\
& & 4\{ (E(p(x)) +  E(\e(x)))  E(u_{(a}) E(q_{(c'}) +  
  (E(p(x')) + E(\e(x')))  E(q_{(a}) E(u_{(c'})\} \C[u_{b)}, u_{d')}]  \\
 & & \tilde{N'}^{(field)}_{ab} \longrightarrow  
\frac{1}{2} \tilde{N}^{'(fluid)}_{ab} 
 \mbox{(relating the corresponding term as:)}
\nonumber\\
& & \{- G_{;p'b} \tG{_;^{p'}_a } -m^2  \tG{_;_a} \tG{_;_b} \} \longrightarrow
\frac{1}{2}\{ E(u_a) E(u_b) \{ \C[\e(x),p(x')] + \C[p(x),p(x')] \}  \} \\
 & &   \tilde{N}^{(field)}  \longrightarrow \frac{1}{2} \tilde{N}^{(fluid)}
 \mbox{(relating the corresponding term as: ) } \nonumber \\
& & \{ \frac{1}{2} \tG{_;_{p'}_q} \tG{_;^{p'}^{q}} +
 \frac{1}{2}  m^2 [  G_{;p'} \tG{_;^{p'}} +  G_{;p} \tG{_;^p} +  m^2 G^2] \}
 \longrightarrow \frac{1}{2} \{ \C[p(x), p(x')] \} \\
 & &   \tilde{N}^{(field)}_{c'd'} \longrightarrow 
\frac{1}{2} \tilde{N}^{(fluid)}_{c'd'} 
 \mbox{(relating the corresponding term as ) } \nonumber \\
& &   2  \tG{_;_{c'}_p} \tG{_;_{d'}^p} 
 \longrightarrow 
\frac{1}{2} \{  E(u_{c'}) E(u_{d'}) \{ \C[p(x), p(x')] + \C[p(x), \e(x')] \}  \}
\eea
\end{widetext}
One can compare this, with the expressions for perfect fluid in 
\cite{seemacorr,seemacqg} and observe the difference. As we have considered the
four-velocity also as a random variable in the present article, 
the expressions with the expectation $E(u_a) $ as well as
covariances of four-velocity vectors appear in the results.
This makes it difficult to obtain reverse one to one correspondences as in
\cite{seemacorr}.  
 The results obtained here, are relevant for the quantum fluids
with the stochastic effects showing up in terms of the kinetic 
as well as bulk variables of the fluid.  Also one can assign thermal
fluctuations in the quantum fluids with this prescription.  We emphasize that
 our scales of interest are with the sub-hydro mesoscopic
range physics, but these expressions may also be used for large scale
hydrodynamic description, if one considers fluctuations w.r.t time 
in a conventional stochastic description rather than the generalized stochastic
description.   Another  interesting possible direction of investigations with
these results can be about quantum potential of scalar fields \cite{qpot},
which is responsible for pressure in the fluid approximation for
dark matter. Hence
it would be interesting to explore and connect the exact mechanism  in this
regard,
of how the fluctuations of the scalar fields characterize the pressure
 fluctuations in the fluid approximation.  
\section{Concluding Remarks }
In this article we have obtained a relation between the two point fluctuations
of a quantum scalar field and that in the hydrodynamic fluid approximation.  
These results indicate that, fluctuations of quantum fields can induce or
can be approximated with  the 
 sub-hydro mesoscopic effects in the fluid description of matter. The 
  "generalized covariances" (or variances) of pressure, energy
 density, heat flux etc describe them in terms of fluid variables.    
  These results can be applicable for the perturbative theory in general
 relativity as the noise or source of perturbations.
 The significance also lies in realising their importance for
compact astrophysical objects which are coupled to (say) thermal fields 
 and are of interest to collapsing clouds, towards critical phases and end
 states of collapse where fluctuations can play a critical role.
Thus our results can be used to analyse properties of dense compact matter
with the relativistic fluid models in strong gravity regions, and to 
study their dynamics at intermediate length scales which become interesting 
with  such a stochastic analysis.
 The extended (point separated) structure and properties given in terms of two
 point statistical covariances of matter fields  is the
 key feature in this article. We have shown the  correspondences 
between the quantum field fluctuations and fluid fluctuations as a 
first principles approach, theoretically. One can progress on these lines of
investigation to explore  more features which connect microscopic effects
in a coarse grained description to see the effects that filter out to
the mesoscopic range. These will characterize the physics at the intermediate
scales in the quantum fluids. The significance of our correspondences lies
in realizing that, these expressions are the starting point for a range of
directions as applications.
 Some of the further questions that arise with this and are open for 
reflections at the very fundamental level in research are, 
\begin{itemize}
\item  How do the quantum field fluctuations leak or tunnel out into 
intermediate  scales in quantum fluids in terms of fluctuations of the
 fluid variables.  Here we have just shown the correspondence between the two 
cases from theoretical considerations, but not addressed how the
physical correspondence is achieved. This can be 
compared with decoherence, where the classical results are obtained from
quantum states, but the phenomena of decoherence is  in itself a different
investigation altogether. 
\item As one may expect, that the stochastic fluctuations averaged out at a
 given
 point in spacetime would be vanishing, we have defined in terms of the
noise kernel two point correlations of the fluctuations, which are 
meaningful. These expressions address not just the equilibrium 
configuration of the
system but also the non-equilibrium configurations can be analysed. Thus one
 can explore various  statistical properties with such a prescription, which are
otherwise inaccessible with the purely quantum and classical  studies.
One of the  main questions that one can ask here is, how does the quantum
fluid behave at intermediate mesoscopic length scales, can these two point 
or higher correlations of the fluctuations inform us of the global  or 
extended nature of the fluid.  What are the length scales in different
types of relativistic fluids which can be correlated in such a fashion 
for mesoscopic studies.
Do these correlations extend to macroscopic range in the fluids, and if yes,
then which interesting new results can be obtained from them for the 
hydrodynamic macroscopic scales. 
\item The quantum fields have a correspondence with fluids is known, we have
shown here how the fluctuations have a correspondence each other. 
 This is  certainly a quantum to classical
correspondence description, as the fluid  fluctuations are
given in terms of classical physical variables.  The question one can ask is, 
 to what extent does the quantum nature of the field fluctutaions tunnel 
through or is retained in the fluid variables fluctuations. Quantum fluids
 are certainly bulk systems, hence they are interesting from the point of veiw
of exploring how quantum nature of matter behaves in the bulk and which 
unique properties show up.  Scalar field models for dark matter 
\cite{qpot,sahni} are
another interesting application for such correspondence and exploring
the fluctuations on our lines of thought can 
have interesting applications for theoretical considerationg in this area. 
\item Thermal and   quantum mechanical fluctuations are widely understood as
the two basic types of  stochastic fluctuations due to thermal effects and
quantum mechanical origin respectively, in physical systems. However  the
 origin of fluctuations in
physical quantities and natural systems can have other causes, like 
in purely mechanical systems coupled to each other, the dynamics or mechanical
 effects can give rise to stochastic vibrations in the system. For example in
 massive gravitating bodies, random oscillations due to any external
or internal mechanical effects can be realized in certain physical 
parameters.   
Whether the correspondence obtained here in terms of fluid variables
can be related to only the thermal fluctuations or we can extend
 this to
non-thermal and purely mechanical effects due to the dynamical motion of fluid
 particles of the relativistic matter under the influence of strong gravity, is
 an interesting query. 
\item Dissipative relativistic fluids are an active area of research 
\cite{nils2} and modelling dissipation in the hydrodynamic approximation is
 non-trivial.
Yet  models of relativistic fluids with dissipation leave provisions
for including fluctuations from the first principles. Though 
fluctuation-dissipation relations have been obtained for relativistic fluids
\cite{kovtun}, there is
still a large scope to construct basic models where the fluid variable
fluctuations can be related with the dissipative effects. One 
can ask the question, if probed through mesoscopic scales in relativistic
fluids, can one define and relate fluctuations in a few of the fluid
 variables, like that of of pressure, energy density and four velocity
present at mesoscopic scales    to dissipative
parameters in the hydrodynamic description.  This calls for including
the theory of generalized fluctuations in relativistic hydrodynamics from the
 first principles, and not just as an ad-hoc assumption.  
\end{itemize}  
Some of the applications oriented directions in research with these developing
ideas are as following, 
\begin{itemize}
\item  The correspondence  of the two point fluctuations given here can be
used in the semiclassical Einstein-Langevin equation \cite{bei2} for 
cosmological spacetime, where the inflaton field can be replaced by the fluid
approximation.  The fluid fluctuations would then represent the thermal states
 as
equivalence of quantum fluctuations in a statistical description. Further on
these lines, one will have to cast the semiclassical Einstein-Langevin 
equation with matter sector represented in terms of fluid equivalence of
the quantum fields. Thus the next step would be to obtain the dissipation 
kernel in the semiclassical Einstein-Langevin equation in terms of the
fluid variables.  
\item  For dense matter fluids  that compose the relativistic stars,  one
can explore the perturbations that  can be induced due to these internal
generalized stochastic fluctuations. It is the cumulative effect of the
 fluctuations  that one would be interested in, to see the perturbations of 
the trajectories of the dense matter  in
 compact stars.  Such ideas \cite{seema1,seema2, seema3} have just begun 
with the proposal and formalism of a classical Einstein-Langevin equation, and
 make way for further scope in asteroseismology. The noise kernel presented
 here, acts as a source for perturbations in relativistic stars, 
and thus its
characterization is specific to the given configuration that
one is interested in exploring.   
\end{itemize}
An interesting direction  can be seen to emerge from this work
 for studying  microscopic structure in matter fields and its connections with
 kinetic theory in curved spacetime \cite{blbook}. Such an endeavour can begin
 by trying
 to consider these generalized fluctuations (including roughness of physical
variables ) of matter fields as  more fundamental  than trying to define 
 particles in a curved spacetime.
 We know that a global definition for particles and that for vaccum  in a curved
spacetime background is not unique. One may then attempt to formulate a
 kinetic theory using the field fluctuations and its generalization as the
 basic entity. 
 This approach may find its way through the four-velocity which 
represents  the kinetic term and its generalized stochastic fluctuations as
 given in this article . 
  With the framework of two point or higher correlations of
  fluctuations of matter fields, a tool to study non-local and
 extended structure of matter in the curved spacetime arises in an 
interesting way. Thus a kinetic theory
of matter in curved spacetime  can be based on these fluctuations
and their interaction, rather than  on the ambiguous localized particles.

\end{document}